\documentstyle[eqsecnum,aps,graphicx]{revtex}

\begin{document}
\draft
\title{S-wave $\pi^0$ Production in pp Collision in a
OBE Model}
\author{ E.Gedalin\thanks{gedal@bgumail.bgu.ac.il},
 A.Moalem\thanks{moalem@bgumail.bgu.ac.il}
 and L.Rasdolskaya\thanks{ljuba@bgumail.bgu.ac.il}}
\address{ Department of Physics, Ben Gurion University, 84105,
Beer Sheva, Israel}
\maketitle
\begin{abstract}
The total cross section for the $pp \rightarrow pp \pi^0$
reaction at energies close to threshold is calculated using
a covariant one-boson-exchange model, where a boson B 
created on one of the incoming protons is converted into
a neutral pion on the second. The amplitudes for 
the conversion processes, $B N \rightarrow N \pi^0$,
are taken to be the sum of  s, u and t-channel pole terms. 
The main contributions to the primary production
amplitude is due to an effective isoscalar $\sigma$ meson pole in 
a t-channel, which is enhanced strongly due to offshellness. 
With this contribution included the model reproduces,
both the scale and energy dependence of the cross section.\\


\end{abstract}
\ \\

\bigskip

\newpage
\section{Introduction}
In a recent study of the $pp \rightarrow pp \pi^0$  reaction 
at energies close to threshold, it is found that the 
angular distributions of the outgoing particles are isotropic, 
in agreement with the assumption that the
reaction proceeds as a $^{33}P_0  \rightarrow{}^{31}S_0 s_0$ 
transition\cite{bondar95,meyer92}.
Several model calculations\cite{koltun66,laget87,miler91,niskanen92} 
of S-wave pion production, which are based on
a single nucleon and a pion rescattering mechanism, 
under estimate the cross section by a factor of 3-5. 
Inspired by a study of $\beta$-decay in nuclei, which indicates
that the axial charge of the nuclear system is 
enhanced by heavy meson exchanges and, by using a simple 
operator form of the NN potential, Lee and Riska\cite{lee93} 
have shown that meson exchange currents could 
explain the scale of the cross section.
Adding to the single nucleon and rescattering terms a Z-graph 
describing various heavy meson exchanges, Horowitz et al.\cite{horowitz94} 
performed similar calculations based on explicit one-boson-exchange 
(OBE) model for the NN interaction as well
as for the evaluation of meson exchange contributions. 

The contribution from the rescattering term depends on the  
off-shell behavior of the $\pi N \rightarrow \pi N$ scattering
amplitude. There are several approaches based on field 
theoretical models which allow such extension to be made but 
the results are model dependent. In the traditional 
phenomenological treatment\cite{koltun66,laget87,miler91,niskanen92}, 
the off mass-shell and on mass-shell amplitudes are of the same 
order of magnitude. Recently, several 
groups\cite{hernandez95,hanhart95,park96,cohen96,sato97}
have shown that the off mass-shell $\pi$N rescattering 
amplitude is enhanced strongly with respect to the on mass-shell amplitude. 
Hernandez and Oset\cite{hernandez95} by applying current 
algebra and PCAC constraints argue that 
this enhancement may bring the calculated cross 
section for the $pp \rightarrow pp \pi^0$  reaction 
into agreement with experiment. However based on 
detailed momentum-space calculations, Hanhart et 
al.\cite{hanhart95} conclude that the enhancement of the 
rescattering amplitude due to offshellness falls a bit too short 
to explain the scale of the $pp \rightarrow pp \pi^0$ cross section. 
Park et al.\cite{park96}, Cohen et al.\cite{cohen96} and Sato 
et al.\cite{sato97}
have applied a chiral perturbation theory ($\chi$PT), including
chiral order 0 and 1 Lagrangian terms. They have shown that the 
off mass-shell $\pi$N elastic scattering amplitude is enhanced 
considerably but has an opposite sign with respect to 
the on mass-shell amplitude. Because of this 
difference in sign, the rescattering term and the Born term 
contributing to the $pp \rightarrow pp \pi^0$  reaction 
interfere destructively, making the theoretical
cross sections much smaller than experimental values and thus
suggesting a significant role for heavy meson exchanges in $\pi^0$ 
production.

In the present work we report on a covariant OBE model 
calculations based on the mechanism depicted in Fig. 1, 
where a virtual boson 
B (B=$\pi$, $\sigma$, $\eta$, $\rho$, $\omega$, $\delta$ ...) 
created on one of the incoming nucleons, is converted into 
a $\pi^0$ meson on the second via a  $BN \rightarrow \pi^0 N$ 
conversion process. 
The half off mass-shell amplitudes for the conversion processes, 
hereafter denoted by $T_{BN  \rightarrow \pi^0 N}$, are
taken as the sum of three terms corresponding to the s, u and 
t-channels displayed in Fig. 2. 
One important aspect of this mechanism is that at 
the $\pi^0$ production threshold, the transferred 
4-momentum is space-like,  $q^2  = -3.3 fm^{-2}$. 
This is very much the same kinematic as occurring in $\pi^0$
electroproduction through vector meson exchanges, 
and we apply a formalism similar to the one applied 
for electroproduction amplitudes\cite{pilkuhn} calculations.
The formalism to be applied is consistent with the OBE 
picture of the
nuclear force and accounts for relativistic effects, 
crossing symmetry, energy 
dependence and nonlocality of the hadronic interactions. 

We start with the assumption that the pion production in 
$pp \rightarrow pp \pi^0$
can be described by the graphs of Fig. 3. 
The diagrams 3a and 3b represent  nucleon pole terms
where the neutral pion is produced on a nucleon 
line. The diagram 3c depicts a pion production occurring on an
internal meson line.  By applying s-channel unitarity to NN elastic 
scattering in the energy region below two-pion production threshold, 
it can be shown that mechanisms as such should contribute 
to the  production process also\cite{pena96}. Sch$\ddot u$tz et 
al.\cite{schutz94} reported on $\sigma$ and $\rho$ meson t pole  
contributions to $\pi$N scattering. Their model was based on
correlated $2\pi$ exchange and constrained by using quasiempirical
information about the $N{\bar N} \rightarrow \pi \pi$ amplitudes.
Hanhart et al.\cite{hanhart95} used the T-matrix obtained in this
model to calculate the rescattering term for $pp \rightarrow pp \pi^0$.
More recently, van Kolck et al.\cite{kolck96} considered other short
range meson exchange mechanisms like $\rho \rightarrow \omega \pi$
and $\delta \rightarrow \eta \pi^0$. To keep our model calculations
consistent with the OBE picture of Machleidt\cite{machleidt89} we limit
our discussions to contributions from $\sigma$, $\delta$ and $\rho$
poles. In variance with previous calculations\cite{schutz94,kolck96}
we write these vertices in a more general form  and determine the relevant
coupling constants by applying the Adler's consistency conditions and/or
fitting to some observables.
We shall demonstrate below that a $\sigma$-meson t pole term
which accounts $effectively$ for isoscalar-scalar two-pion
exchanges contributes significantly to $\pi^0$ production.
With this contribution included the model explains  
the cross-section data for the $pp \rightarrow pp \pi^0$
reaction at threshold.

The present article is organized as follows. In Sect. II we present details 
of the formalism and model parameters. The evaluation of the 
$T_{\pi^0 N \rightarrow \pi^0 N}$, $T_{\eta p \rightarrow \pi^0 p}$
and $T_{\rho p \rightarrow \pi^0 p}$ 
conversion amplitudes require knowledge of the 
$\sigma \pi \pi$, $\delta \eta \pi$ and $\rho \omega \pi$ vertices. 
These are deduced in subsections B-D.  In Sect. III
we write the S-wave production amplitude in a form suitable for 
numerical calculations. Model predictions are given in Sect. IV.
At energies close to threshold, final
state interactions (FSI) influences the energy dependence of the 
calculated cross section\cite{hep95}. These and initial state 
interactions (ISI) corrections are introduced in Sect. V
where comparison with data is to be made. We conclude and summarize 
in Sect. VI. 

\section{The model}
We use the following Lagrangian interaction: 
\begin{eqnarray}
L & = & \frac {f_{\pi NN}}{m_{\pi}} \bar{N} \gamma^5\gamma^{\mu} 
\partial_{\mu} {\bf \pi}{\bf \tau} N +
\frac {f_{\eta NN}}{m_{\eta}} \bar{N} \gamma^5\gamma^{\mu}
\partial_{\mu} \eta N + 
g_{\sigma NN} \bar{N} {\bf \sigma} N  +  \nonumber \\
  &   & g_{\rho NN}\bar{N}\left(  \gamma^{\mu} +
        \frac {\kappa_V}{2M}\sigma^{\mu \nu} \partial_{\nu}\right)
{\bf \tau} {\bf \rho}_{\mu} N +
g_{\omega NN} \bar{N} \gamma^{\mu} {\bf \omega}_{\mu} N +
g_{\delta NN} \bar{N}{\bf \tau} {\bf \delta}  N~,
\label{eq:1}
\end{eqnarray}
with obvious notations. This expression includes terms 
identical to the ones
used by Machleidt et al.\cite{machleidt89} to fit NN elastic 
scattering data in the energy region 0-420 MeV. Here as in 
Ref. \cite{machleidt89} pseudovector couplings are assumed for 
the pseudoscalar mesons and the $\omega$ tensor coupling 
is taken to be zero. All of the coupling constants, meson 
masses and cut off parameters  are  taken from Table A.2 of 
Ref.\cite{machleidt89}; their potential C parameter set.

To calculate the transition amplitude we assume that the reaction 
is dominated by the mechanism depicted in Fig. 1 and write 
the primary production amplitude as, 
\begin{equation}
M^{(in)} = \sum_{B}^{} \left[ T_{BN  \rightarrow \pi^0 N}
 (p_4,k;p_2,q) G_B (q_{})
S_{BNN} (p_3,p_1)\right] + 
[1 \leftrightarrow 2 ; 3 \leftrightarrow 4]~.
\label{eq:2}
\end{equation} 
Here $p_i$, $q$ and $k$ are 4-momenta of the i-th nucleon, 
the exchanged boson and the outgoing pion. The sum runs over 
all possible B bosons that may contribute to the process. 
To be consistent with the OBE picture of the NN force\cite{machleidt89}, 
we include exchange contributions from all of the   
$ \pi, \eta, \sigma , \rho , \omega$ and $ \delta$ mesons.
The bracket $[1 \leftrightarrow 2 ; 3 \leftrightarrow 4]$ 
stands for a similar sum with the $p_1$, $p_3$ 
and $p_2$, $p_4$ momenta interchanged. In Eqn. \ref{eq:2} , 
$T_{BN  \rightarrow \pi^0 N}$ represents the conversion amplitude 
corresponding to the $BN  \rightarrow \pi^0 N$ process. The quantities
$G_B (q)$ and $S_{BNN} (p_3,p_1)$ are the propagator and source 
function of the meson exchanged, respectively. Throughout this 
work we use covariant expressions for the meson propagators 
and form factors as defined in Ref.\cite{machleidt89}. The source 
functions for scalar, pseudoscalar and vector mesons are
\begin{eqnarray}
 & & S_{SNN} (p_1,p_3) = \bar {u} (p_3)\  I\  u (p_1)\  F_S (q)~,\\
 & & S_{PNN} (p_1,p_3) = 
\bar {u} (p_3)\  \gamma^5\  I\  u (p_1)\  F_P (q)~,\\
 & & S^{\mu}_{VNN} (p_1,p_3) = 
\bar {u} (p_3)\left[ \gamma^{\mu} F^{(1)}_V (q^2_{13})
 +i \sigma^{\mu \nu} q_{\nu} F^{(2)}_V (q^2_{13}) 
 \right] I u (p_1)~,
\label{eq:3}
\end{eqnarray} 
where $u(p)$ stands for a nucleon Dirac spinor; $I$ is the appropriate 
isospin operator and $p_3 = p_1 - q$ is the final nucleon momentum. 
The functions $F_S(q)$ and $F_P(q)$ are source form factors
for scalar and pseudoscalar mesons. For vector mesons there 
are two such
quantities  $F_V^{(1)}$ and $F_V^{(2)}$ representing vector and tensor 
form factors, the  analogous of the nucleon 
electromagnetic form factors. In the calculations to be presented 
below all source form factors are taken in the form\cite{machleidt89}, 
\begin{equation}
F_B (q^2) = g_{BNN} f_B(q);~~~~~~f_B (q) = 
 \frac {\Lambda_B^2 - m_B^2}{\Lambda_B^2 - q^2}~.
\label{eq:4}
\end{equation}
Albeit, the propagators and source functions are rather well 
determined from fitting NN scattering data \cite{machleidt89}.
Thus to a large extent, the model success in explaining 
cross section data for the $pp \rightarrow pp \pi^0$ reaction 
depends on how well the conversion amplitudes 
$T_{B N \rightarrow \pi^0 N}$ are calculated. The relative 
importance of the various exchange contributions depend upon the
off mass shell behavior of these amplitudes.
For example, though very small on 
mass-shell, the amplitude for $\pi^0 p  \rightarrow \pi^0 p$  
process is strongly enhanced off mass-shell \cite{hernandez95,hanhart95} 
giving rise to a dominant contribution to the production rate.

\subsection{Nucleon pole s and u channel amplitudes}
In this subsection we write expressions for s and u-channel
nucleon pole terms 
(diagrams 2a and 2b). These are common to all of the 
various meson exchanges. We call  p and p' the momenta of
incoming and outgoing nucleon, q and k those of the boson B and the pion
produced, respectively. Let $s=(p'+k)^2$ be the total
energy and
\begin{equation}
\Delta_N (x) = \frac {i}{x^2 -M^2 +i \epsilon}~.
\label{eq:6}
\end{equation}
Using the vertices of the Lagrangian, Eqn. 1, 
and the usual Feynman rules the contributions from diagram
2a with B designating a scalar (S), a pseudoscalar (P) and 
a vector meson (V) are respectively, 
\begin{eqnarray}
 & & T_S^{(s)} = - \frac {g_{SNN}f_{\pi NN}}{m_{\pi}}
 f_{S}(q) f_{\pi}(k)\bar {u}(p') \gamma^5 k\! \! \! /
     \frac {p'\! \! \! / + k\! \! \! /  + M}{(p'+k)^2 - M^2} u(p)~,\\
 & & T_P^{(s)} = i \frac {f_{PNN}f_{\pi NN}}{m_{P}m_{\pi}} 
      f_{P}(q) f_{\pi}(k) \bar {u}(p') \gamma^5 k\! \! \! /
     \frac {p'\! \! \! / + k\! \! \! /  + M}{(p'+k)^2 - M^2} 
     \gamma^5 q\! \! \! / u(p)~,\\
 & & T_V^{(s)}= - \frac {g_{VNN} f_{\pi NN}}{m_{\pi}} 
     f_{V}(q) f_{\pi}(k) \bar {u}(p') \gamma^5 k\! \! \! /
     \frac {p'\! \! \! / + k\! \! \! /  + M}{(p'+k)^2 -M^2} 
  \left[ \gamma^{\mu} + \frac {\kappa_V}{2M} \sigma^{\mu \nu} 
q_{\nu}\right]
  u(p)~,
\label{eq:7}
\end{eqnarray}
where $a\! \! \! / = a^{\mu}q_{\mu}$. After some algebra one obtains,
\begin{eqnarray}
 & & T_S^{(s)} = \nonumber\\
 & &i \frac {g_{SNN}f_{\pi NN}}{m_{\pi}}
       f_{S}(q) f_{\pi}(k) 
     \bar {u}(p') \gamma^5 \left[ (s -M^2) + 2M k\! \! \! / \right] u(p)
     \Delta_N (s)~,
\label{sspole}\\
 & & T_P^{(s)} = (2M)^2 \frac {f_{PNN}f_{\pi NN}}{m_P m_{\pi}}
     f_{P}(q) f_{\pi}(k) \nonumber\\
 & &    \bar {u}(p') \left[-\left(\frac {s-M^2}{2M}\right) +  k\! \! \! /
     \left(\frac {s-M^2 + 4M^2}{4M^2}\right)\right]\Delta_N(s)u(p),\\
\label{pspole}
 & & T_V^{(s)}= i \frac {g_{VNN} f_{\pi NN}}{m_{\pi}}\Delta_N(s) 
     f_{V}(q) f_{\pi}(k)
     \bar {u}(p') \gamma^5 \nonumber \\
 & & \left[\gamma^{\mu} (s-M^2) + 4k^{\mu} -2M\gamma^{\mu}
     k\! \! \! / \right] \left[ 1 + \left(\frac {\kappa_V}{2M}\right)
     q\! \! \! / \right]
     u(p)~.
\label{vspole}
\end{eqnarray}
The contributions from the u-channel (diagram 2b) are obtained easily by 
replacing the total energy s in Eqns. 11-13 with $u = (p-k)^2$. 
Likewise, for a vector meson the u-channel pole term is
\begin{equation}
T_V^{(u)} = - i \frac {g_{VNN} f_{\pi NN}}{m_{\pi}}\Delta_N(s)
     f_{V}(q) f_{\pi}(k) 
     \bar {u}(p') \gamma^5 \gamma^{\mu} \left( 1 - \frac {\kappa_V}{2M}
     q\! \! \! /\right) \left[u-M^2 + 2M
     k\! \! \! /\right] u(p)~.
\label{eq:9}
\end{equation}

\subsection{Evaluation of $T_{\pi^0 N \rightarrow \pi^0 N}$}
We now turn to calculate the amplitude 
$T_{\pi^0 N \rightarrow \pi^0 N}$. In this case in addition to s and
u-channel nucleon pole terms, a neutral pion
can be formed on an internal $\sigma$ meson line.  These terms are 
displayed graphically in Fig. 4. The contribution from graph 4c depends
upon the $\sigma \pi \pi$ meson vertex. There is no information about
this coupling from traditional analyses of NN elastic 
scattering\cite{machleidt89}. In order to 
determine this vertex we apply some physics beyond the OBE
picture of the NN interactions. Quite generally we 
write an effective $\sigma\pi\pi$ vertex in the form,
\begin{equation}
V_{\sigma \pi \pi} (k,q) = m_{\sigma}\left( g_{0\sigma} +
\frac {q^2 + k^2}{m_{\sigma}^2} g_{1\sigma} +
\frac {(q - k)^2}{m_{\sigma}^2} g_{2\sigma}\right)~,
\label{sigver}
\end{equation}
where $g_{0\sigma}, g_{1\sigma}, g_{2\sigma}$ are (as yet unknown) 
constants. This is a slightly generalized Weinberg-type low
energy expansion\cite{weinberg79} allowing for three rather than one
different constants.
Taking the appropriate s and u-channel 
contribution from
Eqn. 12 and adding the contribution from a $\sigma$ pole in a 
t-channel one obtains,
\begin{eqnarray}
 & & T_{\pi^0 p \rightarrow \pi^0 p} =
  i \left( \frac {2Mf_{\pi NN}}{m_{\pi}}\right)^2 
  f_{\pi}(q) f_{\pi}(k){\bar u}(p')
  \left[k\! \! \! / \left( \frac {1}{s -M^2 } - \frac {1}{u - M^2} 
  \right) - \frac {1}{M}\right]  u(p) +~\nonumber \\
 & & i g_{\sigma NN} 
 \frac {  f_{\sigma}(q - k) }{(q - k)^2 -m_{\sigma}^2} 
 V_{\sigma \pi\pi} {\bar u}(p') u(p)~.
\label{pnpn}
\end{eqnarray}
Now, in order to determine the vertex constants $g_{0\sigma}, 
g_{1\sigma}$ and $g_{2\sigma}$ we require that the conversion 
amplitude obeys the three Adler's consistency 
conditions\cite{alfaro,adler65}. These are
\begin{eqnarray}
 & & T_{\pi^0 p \rightarrow \pi^0 p} ( k = q = 0 ) = -i 
             \frac {\sigma_{\pi N}(0)}{F_{\pi}^2}~,\\
 & & T_{\pi^0 p \rightarrow \pi^0 p}(k=0, q^2 = m_{\pi}^2) = 0~,\\
 & & T_{\pi^0 p \rightarrow \pi^0 p}(k^2 = m_{\pi}^2 , q = 0) = 0~, 
\end{eqnarray}
where $\sigma_{\pi N} (0)$ is the well known $\pi N$ 
$\sigma$-term\cite{bernard95} and $F_{\pi}$ the pion radiative 
decay constant. These three conditions yield only two relations 
amongst the constants, which are
\begin{eqnarray}
 & & g_{0\sigma} = \frac  {\sigma_{\pi N}(0) m_{\sigma}}
                    {F_{\pi}^2 g_{\sigma NN} f_{\sigma}(0)}~,\label{r:1}\\
 & & g_{0\sigma} = \left(\frac {m_{\pi}}{m_{\sigma}}\right)^2
                   \left( g_{1\sigma} + g_{2\sigma}\right)~.
   \label{r:2}
\end{eqnarray}
Furthermore, we recall that the on mass shell $\pi^0$p S-wave elastic 
scattering amplitude is parameterized according to\cite{koch86,ericson},
\begin{equation}
F = a^+ + b^+ {\bf k}\cdot{\bf k}~,
\label{pinampl}
\end{equation}  
with $a^+$ being the isospin even $\pi^0$p scattering length.
To obtain a third relation we may require that the amplitude Eqn. 17
reproduces $F$ at ${\bf k} = 0$, i.e.,
\begin{equation}
T_{\pi^0 p \rightarrow \pi^0 p}(k^2 = m_{\pi}^2 ,{\bf k} = 0 ; 
q^2 = m_{\pi}^2 , {\bf q}=0) = i 4\pi \left( 1 + \frac {m_{\pi}}{M}
\right) a^+~. \\
\label{r:3}
\end{equation}
This yields,
\begin{equation}
 g_{0\sigma} + 2 (\frac {m_{\pi}}{m_{\sigma}})^2 g_{1\sigma}= 
\frac {-1}{g_{\sigma NN}}\left[
 4 \pi \left(1 + \frac {m_{\pi}}{M}\right) m_{\sigma} a^+
- f_{\pi NN}^2 \left(\frac {m_{\sigma}}{M}\right)
               \left(\frac {4 M^2}{4 M^2 - m_{\pi}^2}\right)\right]  ~.
\label{r:3}
\end{equation}
We now may use $\sigma_{\pi N} (0)$, $F_{\pi}$, $a^+$ and 
$g_{\sigma NN}$ as input to evaluate the vertex constants. The 
value of the quantity $\sigma_{\pi N} (0)$ is related to matrix
elements of the operator $m_q q {\bar q}$ in the proton, 
where $m_q$ stands for
the mass of the proton quark constituents.
This quantity can be calculated from the baryon 
spectrum\cite{gasser91}. To leading orders in the quark masses 
one finds to order ${\it O} (m_q^{3/2})$ that
$\sigma_{\pi N} (0) = 26$ MeV. Including ${\it O} (m_q^{3/2})$ 
contributions and estimate ${\it O} (m_q^{2})$ ones yields,
\begin{equation}
\sigma_{\pi N} (0) = \frac {35 \pm 5}{1-y} ~MeV~,
\label{sigma} 
\end{equation}
with $y$ being a measure of the strange quark content in the proton.
An even higher a value is obtained from the isospin even $\pi$N 
scattering amplitude at the Cheng-Dashen point,  which can be determined
experimentally using the low-energy theorem of current algebra.
This gives\cite{gasser91} a value $\sigma_{\pi N} (0) = 45 \pm 8 ~MeV$. 
Equation \ref{sigma} suggests a value $\sigma_{\pi N} (0) = 35 ~MeV$ 
for $y=0$, in keeping with the OZI rule\cite{ozi}.
With this value and taking  a radiative
decay constant $F_{\pi} = 93.5$ MeV \cite{pdg}; isospin even 
$\pi^0$p scattering length $a^+ = (-0.010 \pm 0.003) m_{\pi}^{-1}$
\cite{koch86,ericson}, and the $\sigma$NN coupling $g_{\sigma NN}^2/4\pi 
= 8.03$\cite{machleidt89} 
we can now solve Eqns. \ref{r:1}, \ref{r:2}, \ref{r:3} to extract
the values,
\begin{eqnarray}
 g_{0 \sigma} = 0.22 \pm 0.04 ,~~~ & g_{1 \sigma} = -0.82 \pm 0.42,~~~
 & g_{2 \sigma} = -2.62 \pm 0.7~. 
\end{eqnarray}
With these constants Eqns. \ref{pnpn} predicts 
$b^+ = (-0.068 \pm 0.015) m_{\pi}^{-3}$ 
in close agreement with the experimental value 
$b^+_{exp} = -(0.088 \pm 0.014)m_{\pi}^{-3}$ 
quoted by Koch\cite{koch86}. Thus the amplitude Eqn. \ref{pnpn} 
on the mass shell reproduces the correct ${\bf k}\cdot{\bf k}$
dependence  of the $\pi^0$p elastic scattering amplitude, 
Eqn. \ref{pinampl}, also. 

It is difficult to ascertain that the amplitude 
Eqn. \ref{pnpn} does
reproduce the off mass shell behavior correctly. However we 
may confront Eqn. \ref{pnpn} with predictions from other approaches. 
For example, taking the residue of the amplitude, Eqn. \ref{pnpn}, 
at the $\sigma$ pole $(q-k)^2 = m_{\sigma}^2$, one obtains
an effective $\sigma \pi \pi$ coupling
\begin{eqnarray}
g_{\sigma \pi \pi}^{eff} = V_{\sigma \pi \pi} (q^2=m_{\pi}^2, 
k^2 = m_{\pi}^2, 
(q-k)^2 =m_{\sigma}^2) = (2.5 \pm 0.9) m_{\sigma}~,
\end{eqnarray}
a value to be compared with the well known estimate from
soft pion physics\cite{alfaro} 
\begin{eqnarray}
V_{\sigma \pi \pi} (m_{\pi}^2, m_{\pi}^2, m_{\sigma}^2) = m_{\sigma}
\frac {m_{\sigma}}{ 2 F_{\pi}} \left( 
               1 - \frac {m_{\pi}^2}{m_{\sigma}^2} \right)
= (2.8 ) m_{\sigma}~.
\end{eqnarray}

Two-pion loops contribute to the $\pi \pi$ elastic scattering 
amplitude  and appear as corrections to the leading contact term
\cite{alfaro,gasser95}. As a further check we may evaluate these
corrections in terms of an effective $\sigma$-meson exchange in
s, u and t-channels. By doing so it is straightforward to show 
that these corrections are 
\begin{eqnarray}
 & & T_{\pi \pi \rightarrow \pi \pi}^{loop} = 
\frac {2}{m_{\sigma}^2 -s} V_{\sigma \pi \pi}^2 (m_{\pi}^2, m_{\pi}^2, 
s)\delta^{ab}\delta^{cd} +
\frac {2}{m_{\sigma}^2 -u} V_{\sigma \pi \pi}^2 (m_{\pi}^2, 
m_{\pi}^2, u)\delta^{ad}\delta^{bc} +  \nonumber \\
 & & ~~~~~~~~~~~~~~~~~~~~~~~~~~
\frac {2}{m_{\sigma}^2 -t} V_{\sigma \pi \pi}^2 
(m_{\pi}^2, m_{\pi}^2, t)\delta^{ac}\delta^{bd}~,
\label{pipil} 
\end{eqnarray}
where $\delta^{i,j}$ is the Kroneker $\delta$ and with $i,j$ 
being pion isospin indices. The factor of 2 is due to the symmetry
of the two $\sigma \pi \pi$ vertices. 
At threshold this expression amounts to about $20\%$ of the 
contact term, in good agreement with the value of $25\%$ 
reported by Gasser\cite{gasser95} from improved low energy
theorems. 

Now that the $\sigma \pi\pi$ vertex parameters are well defined we 
may examine offshellness effects.  To this aim we evaluate 
$T_{\pi^0 p \rightarrow \pi^0 p}$ at the production threshold
of $pp \rightarrow pp \pi^0$ and compare with its on mass 
shell value. 
When both pion legs are on the mass shell the last term in 
Eqn.\ref{pnpn} reduces to,
\begin{eqnarray}
 & & T_{\pi^0 p \rightarrow \pi^0 p}^{(t)} 
       (k^2 = m_{\pi}^2;q^2 = m_{\pi}^2) 
= - i \frac { g_{\sigma NN} }{m_{\sigma}}
\left[  g_{0\sigma} + 2 \left( \frac {m_{\pi}}{m_{\sigma}}\right)^2
g_{1\sigma} \right]  f_{\sigma}(0)\nonumber \\
 & & ~~~~~~~~~~~~~~~~~~~~~~~~~~~~~~~~~~~~~\approx -i 0.11 
 \frac { g_{\sigma NN} }{m_{\sigma}}~.
\label{eqn:27}
\end{eqnarray}
At threshold of the $pp \rightarrow pp \pi^0$ 
reaction, the momentum squared of the $\sigma$ meson is 
$(q - k)^2 =(p'- p)^2 \approx -M m_{\pi} $ and the off mass
shell t pole term in  Eqn. \ref{pnpn} becomes,
\begin{eqnarray}
 & & T_{\pi^0 p \rightarrow \pi^0 p}^{(t)} = - i g_{\sigma NN} 
\left(\frac {m_{\sigma}}{m_{\sigma}^2 + M m_{\pi}}\right)
 f_{\sigma}(-M m_{\pi}) \nonumber \\
 & & ~~~~~~~~~~~~~~~\left[  g_{0\sigma} + 
\left( \frac {m_{\pi}}{m_{\sigma}}\right)^2
g_{1\sigma} - \frac {M m_{\pi}}{m_{\sigma}^2}
\left( g_{1\sigma} + g_{2\sigma}\right)\right]~\nonumber\\
 & & ~~~~~~~~~~~~~~~\approx -i 1.0 
 \frac { g_{\sigma NN} }{m_\sigma} ~,
\end{eqnarray}
which is a factor $\approx 9$ larger compared to the on 
mass shell value  Eqn. \ref {eqn:27}.
In Fig. 5 the amplitude $T_{\pi^0 p \rightarrow \pi^0 p}$ is drawn 
as a solid line $vs.$ $q^2$.
The contributions from the s and u channel nucleon pole and 
$\sigma$-meson pole terms are drawn as dashed and dot-dashed 
curves, respectively. 
Off mass shell behavior of
the s and u terms in Eqn. \ref{pnpn} is given by the pion form factor,
$f_{\pi} (q)$, the nucleon propagators and the momentum dependence 
of the $\pi NN $ vertex. That of the t-channel
is affected by the $\sigma$ meson form factor $f_{\sigma} (q-k)$ and
the $V_{\sigma \pi \pi}$ vertex function, Eqn. \ref{sigver}. Note that
on  mass shell both of these contributions are small,
opposite in signs and cancel to large extent. As $q^2$
and $(q - k)^2$ become more negative both terms
become negative and therefore add constructively to the 
conversion amplitude, giving rise to strongly enhanced off mass
shell amplitude. We expect then that $\pi$ exchange plays 
an important role in the $\pi^0$ production process.

\subsection{The $T_{\eta p \rightarrow \pi^0 p}$ amplitude }
The production of a neutral pion can occur also on an internal 
$\delta$-meson line. To evaluate the  amplitude for the conversion 
process $\eta p \rightarrow \pi^0 p$ we apply a similar procedure
as above. The $\eta$ meson couples to the nucleon isobar $N^* (1535
~MeV)$ strongly\cite{gedalin96} and the main contribution to the 
conversion amplitude for $\eta p \rightarrow \pi^0 p$ is due to 
s and u isobar pole terms (see Fig. 6), hereafter we refer to
as the resonance contribution. The other terms (graphs 6c-6e) 
furnish a background term.
A t-channel would involve a vertex with
$\delta$, $\eta$ and $\pi$ legs defined to be,
\begin{equation}
V_{\delta \eta \pi}(k,q) = m_{\delta}\left[  g_{0\delta} + 
    \frac{ k^2}{m_{\delta}^2} g_{1\delta} +
    \frac{ q^2}{m_{\delta}^2} g_{2\delta} +
    \frac{ (k-q)^2}{m_{\delta}^2} g_{3\delta} \right]~.
\label{vr:dep}
\end{equation}
Here, to account for the fact that the three legs are 
different the vertex is described in terms of four constants .
Taking the sum of all the graphs in Fig. 6, we may write,
\begin{eqnarray}
 & & T_{\eta p \rightarrow \pi^0 p} = \nonumber\\
 & & -i g_{\eta NN^*} g_{\pi NN^*} f_{\eta} (q) f_{\pi} (k) 
\bar {u} (p') \left[ \frac {1}{M_R - \sqrt {s} + i \Gamma /2} + 
\frac {1}{M_R - \sqrt {u} + i \Gamma /2}\right] u(p)\nonumber \\
 & & i\frac {2M f_{\pi NN}}{m_{\pi}} \frac {2M f_{\eta NN}}{m_{\eta}}
     f_{\eta}(q) f_{\pi}(k)
{\bar u} (p') \left[k\! \! \! / \left(\frac {1}{s-M^2} - 
\frac {1}{u-M^2}\right) - \frac {1}{M}\right] u(p) + \nonumber \\
 & & i g_{\delta NN} \frac {f_{\delta}(q - k)} {(q - k)^2 - m_{\delta}^2}
V_{\delta \eta \pi}(k,q)
{\bar u}(p') u(p)~.
\label{vr:dep}
\end{eqnarray}
Here $M_R = 1535 ~MeV$ and $\Gamma = 175 ~MeV$ stand for the mass 
and width of the isobar resonance, and the coupling constants 
are $g_{\eta NN^*} = 2.2, g_{\pi NN^*} = 0.8$ \cite{gedalin96}.

We now apply the  Adler's consistency conditions to the background term
in Eqn. 33 to write the following relations 
amongst the unknown $\delta\eta\pi$ vertex constants (see Eqns. 17-19), 
\begin{eqnarray}
 & & g_{0\delta} = \frac {m_{\delta}}{g_{\delta NN}f_\delta (0)}
\frac {\sigma_{\pi N \rightarrow \eta N} (0)}{F_{\pi} F_{\eta}}~,
\label{eq:eaa} \\
 & & g_{0\delta} = -\left(\frac {m_{\eta}}{m_{\delta}}\right)^2
         \left(g_{2\delta} + g_{3\delta}\right)~,
\label{eq:eab} \\
 & & g_{0\delta} = -\left (\frac {m_{\pi}}{m_{\delta}}\right)^2
         \left(g_{1\delta} + g_{3\delta}\right)~,
\label{eq:eac} 
\end{eqnarray}
where $F_{\eta}$ is the $\eta$ radiative constant; in the limit of 
an exact SU(3) symmetry $F_{\eta} = F_{\pi}$. Likewise, the quantity 
$\sigma_{\pi N \rightarrow \eta N}$ is a $\sigma$-term, 
a quantity related to matrix 
elements of  various quark mass $q\bar q$ in the nucleon,
\begin{equation}
\sigma_{\pi N \rightarrow \eta N} (t) =
\frac {\hat m}{2} \langle p' |{\bar u}u + {\bar d}d |p\rangle~,
\end{equation}
where $\hat m = (m_u + m_d)/2$ is the average of the u and d 
quark masses and
$t = (p'-p)^2$ the transferred momentum  squared. This 
term can be deduced from the kaon-nucleon (KN) 
$\sigma$-terms  and the strange quark mass through\cite{bernard95}
\begin{equation}
\sigma_{\pi N \rightarrow \eta N} (0) = 
\left( \frac {\hat m}{{\hat m} + m_s} \right) 
\left[ \sigma_{KN}^{(1)}(0) - \sigma_{KN}^{(2)} (0)\right]~.
\end{equation}
Taking the quark masses as $m_u = (5 \pm 2 )$ MeV,
$m_d = (9 \pm 3) $ MeV, $m_s = (175 \pm 55 )$ MeV, and the KN 
$\sigma$-term\cite{bernard95} $\sigma_{KN}^{(1)} (0) \simeq (200 \pm 50)$ MeV ,
$\sigma$-term $\sigma_{KN}^{(2)} (0) \simeq (140 \pm 40)$ MeV ,
one obtains,
\begin{equation}
\sigma_{\pi N \rightarrow \eta N} (0) \simeq (2 \pm 1.3) MeV~.
\end{equation}
Now the partial decay width of the $\delta$ meson into a $\pi \eta$ 
pair is\cite{pdg} $\Gamma_{\pi \eta} = 57 \pm 11 \ MeV$. By calculating this
width using the expression Eqn. 32, one obtains a fourth relation
amongst the $\delta \eta \pi$ vertex  constants,
\begin{equation}
g_{0\delta} + \left(\frac {m_{\pi}^2}{m_{\delta}^2}\right) ~g_{1\delta} + 
\left(\frac {m_{\eta}^2}{m_{\delta}^2}\right) ~g_{2\delta} + g_{3\delta}
= \sqrt {\frac {\Gamma 8\pi}{|k|}}
\end{equation}
Resolving this last relation and the Adler's conditions, Eqns. 
\ref{eq:eaa} - \ref{eq:eac}, gives
\begin{equation}
g_{0\delta} = 0.05\pm 0.035 ~,~~~g_{1\delta} = -5.95\pm 1.5 ~,~~~
g_{2\delta} = -3.46\pm 0.1~,~~~g_{3\delta} = 3.30\pm 0.2~. 
\end{equation}
With these constants one finds that at threshold  for the 
$pp \rightarrow pp \pi^0$ reaction, the 
conversion amplitude is $T_{\eta p \rightarrow \pi^0 p} \approx 0.98 \ fm$.
The t-pole contribution amounts to only 
$T_{\eta p \rightarrow \pi^0 p}^{(t)} \approx 0.04 \ fm$
We may thus conclude that a 
$\delta$ meson  pole in a t-channel contributes very little  
to $T_{\eta p \rightarrow \pi^0 p}$ and should play a minor role 
in the $\pi^0$ production process. 

\subsection{The $T_{\rho p \rightarrow \pi^0 p}$ amplitude }
We follow van Kolck et al.\cite{kolck96} and limit the $\rho \omega \pi$
vertex to the form,
\begin{equation}
V_{\rho \omega \pi}(k,q) = -\frac {g_{\rho \omega \pi}}{m_{\omega}}
\epsilon_{\mu \nu \lambda \delta} q^{\mu}k^{\delta}{\bf \rho}^{\nu}
{\omega}^{\lambda}{\bf \pi}~,
\end{equation}
where the coupling constant $g_{\rho \omega \pi}$ 
$\approx -10$, a value fairly well established. 
With this expression, the contribution from a $\rho \omega$ 
exchange mechanism to the $T_{\rho N \rightarrow \pi^0 N}$ is
\begin{eqnarray}
 & & T_{\rho N \rightarrow \pi^0 N}^{(t)} = ig_{\rho \omega \pi}
g_{\omega NN}  f_{\omega}(q-k)
 \frac {k^0}{m_{\omega}}\frac{1}{(q-k)^2 - m_{\omega}^2}\\ \nonumber
 & &\left[\frac{1}{E+M}\left({\bf q\times p}+i{\bf p q\cdot \sigma}-
 i{\bf \sigma q\cdot p}\right) - \frac{1}{E'+M}
 \left({\bf q\times p'}+i{\bf p' q\cdot \sigma}-
 i{\bf \sigma q\cdot p'}\right)\right]~.
\end{eqnarray}
The s and u nucleon pole terms can be calculated from Eqns. 10 and 14.

\section{S-wave amplitudes and cross section }

In this section we write the primary production amplitude in 
a form suitable for numerical integration. We call
\begin{equation}
{\bf \Pi}_j = \frac {\bf p_j}{E_j + M}
\end{equation}
where ${\bf p_j}$ and $E_j$ are the three momentum and total 
energy of the j-th nucleon. For the incoming particles in the
center of mass system (CM) ${\bf \Pi}_1 =-{\bf \Pi}_2={\bf \Pi}$. 
The energy available in the CM system is 
\begin{equation}
Q = \sqrt {s} - 2M - m_{\pi}~, 
\end{equation}
where $s = (p_1+p_2)^2$ is the total energy squared.
We shall calculate the amplitudes and cross sections as 
functions of the variable Q. We also define
\begin{eqnarray}
 & & q^2 = -M(m_{\pi} + Q)~,\nonumber \\
 & & a=1+\frac {Q}{m_{\pi}}
\left( 1 + \frac {m_{\pi}}{M}\right)~,\nonumber \\
 & & b=1+\frac {Q}{2M}
\left( 1 + \frac {m_{\pi}}{2M} + \frac {m_{\pi}Q}{4M^2}\right)~,\\
 & & R = \frac {1}{M_R - M - m_{\pi} - Q + i \Gamma /2} +
\frac {1}{M_R - M + m_{\pi} + Q + i \Gamma /2} ~. 
\end{eqnarray}
Following the discussion in the previous section we write the
primary amplitude as
\begin{equation}
M^{(in)} = M_{\pi} + M_{\eta} + M_{\sigma} + M_{\rho} + 
M_{\omega} + M_{\delta}~,
\label{primpa}
\end{equation}
where $M_B$ stands for a partial production amplitude representing the contribution
from the exchange of a boson B. We substitute equivalent two-component 
free spinor matrix elements (Table A6.1 of Ref. \cite{ericson})
in Eqns. \ref{sspole}-\ref{vspole} for s-channels and the appropriate 
ones for u-channels, to write the scalar and vector meson exchange 
amplitudes as,
\begin{equation}
M_S = - i ~\frac {f_{\pi NN}}{m_{\pi}} ~\frac {g_{SNN}^2}{m_S^2 -q^2} 
~f_{S}^2(q) ~\frac {m_{\pi} + 2Q}{M + Q}~\frac{M}{2M + m_{\pi}}~
{\bf \Pi} \cdot ( \sigma_1 - \sigma_2 )~,
\label{sexamp}
\end{equation}
and
\begin{equation}
M_V = i ~\frac {f_{\pi NN}}{m_{\pi}}
       ~\frac {g_{VNN}^2}{m_V^2 - q^2} ~f_{V}^2(q)
\left[ X ~{\bf \Pi} \cdot (\sigma_1-\sigma_2) 
+i Y ~{\bf \Pi} \cdot \sigma_1\times\sigma_2 \right]
\label{vexamp}
\end{equation}
where we have used the notation
\begin{eqnarray}
 & & X=\left[2\left(1 + \frac{3m_{\pi}}{2M}\right) + 
\kappa_V \left(1 + \frac{m_{\pi} + Q}{2M}\right)
 \left(1 - \frac{m_{\pi} + Q}{2(2M + m_{\pi})}\right)\right]
(1 - \kappa_V {\bf \Pi}\cdot {\bf \Pi}) 
 \nonumber\\
 && -~\left[\frac{3m_{\pi} + Q}{M}+ 
 \kappa_V \frac{m_{\pi} + Q}{2M}
 \left(1 - \frac{m_{\pi}}{2(2M + m_{\pi})}\right)\right] 
 (1 + \kappa_V {\bf \Pi}\cdot{\bf \Pi})\\
 & & Y=-2 -2(1 + \kappa_V)\left[\left(\frac{3m_{\pi} +Q}{M}\right) + 
\kappa_V \left(\frac{m_{\pi} + Q}{2M}\right)
 \left(1 - \frac{m_{\pi}}{2(2M + m_{\pi}}\right)\right]~.
\end{eqnarray}
The $\rho \omega$ exchange mechanism contribute to the production
amplitude a term,  
\begin{equation}
M^{(t)}_\rho = 
2\frac {g_{\rho NN}}{m_{\rho}^2 - q^2} ~f_{\rho}^2(q)
2\frac {g_{\omega NN}}{m_{\omega}^2 - q^2} ~f_{\omega}^2(q)
g_{\rho \omega \pi}\frac{m_\pi}{m_\omega}({\bf p\cdot \Pi})
{\bf \Pi\cdot\sigma_1 \times \sigma_2}~.\\
\end{equation}
To avoid double counting we add this to $M_{\rho}$ (but not
to $M_{\omega}$) only.  
For both of the $\pi$ and $\eta$ exchange amplitudes, 
there is a t-channel pole
in addition to s and u-channel nucleon pole terms. The sum of all three 
gives,
\begin{eqnarray}
 & & M_{\pi} = -i f_{\pi NN} \left( \frac {2M}{m_{\pi}}\right)
               \left( \frac {1}{m_{\pi}^2 - q^2}\right)
               f_{\pi}(q) \nonumber \\
 & & \left[ f_{\pi NN}^2 \left(\frac {1} {m_{\pi}}\right) 
               \left( \frac {a}{b} \right) f_{\pi}(q) + 
  g_{\sigma NN} \left(\frac {1} {m_{\sigma}^2 - q^2}\right)
  f_{\sigma}(q)
  ~V_{\sigma \pi \pi}(k^2 = m_{\pi}^2 ; q^2) 
               \right]~,
\label{piexamp}
\end{eqnarray}
\begin{eqnarray}
 & & M_{\eta} = -i f_{\pi NN} \left( \frac {2M}{m_{\eta}}\right)
               \left( \frac {1}{m_{\eta}^2 - q^2}\right)
               f_{\eta}(q) \nonumber \\
 & &           \left[g_{\eta NN^*} g_{\pi NN^*} R + 
               f_{\pi NN} f_{\eta NN} 
\left(\frac {f_{\eta}(q)} {m_{\eta}}\right) 
               \left( \frac {a}{b} \right) + 
  g_{\delta NN} \left(\frac {f_{\delta}(q)} {m_{\delta}^2 - q^2}\right)
  V_{\delta \eta \pi}(k^2 = m_{\pi}^2 ; q^2) 
               \right]~.
\label{etaexamp}
\end{eqnarray}

Finally, the total cross section is calculated from the expression,
\begin{equation}
\sigma = \frac {M^4}{16 (2\pi)^5 \sqrt(s) {\bf p}_1}
\int  \frac {d^3{\bf p}_3} {E_3} \ \frac {d^3{\bf p}_4} {E_4}\ 
\frac {d^3{\bf p}_{\eta}} {E_{\eta}}   
\ | \ Z_{}M^{(in)}\ |^2 \delta {}^4 (p_i -  p_f) \nonumber\\ ,
\end{equation}
where  $Z$ is the three-body FSI correction factor 
of Refs.\cite{hep95} to be specified below.

\section{Predictions and comparison with data }
We apply now the model presented in the previous sections to 
calculate the total cross section for the $pp \rightarrow pp \pi^0$
reaction at energies close to threshold. We first consider the 
relative importance of the various exchange contributions. To
this aim we draw in Fig. 7  the primary production amplitude 
$M^{(in)}$ along with the partial exchange amplitudes of Eqns. 46-51, 
vs. the energy available in the center of mass (CM) system. 
The main contribution is due to pion exchange with ratios 
$M_{\pi} : M_{\rho} :  M_{\omega} : M_{\sigma} : M_{\eta} : M_{\delta} 
\approx 138 : 42 :  8 : 6 : 5 : 0.6 $. Next  important to 
the pion is contribution from the $\rho$ 
meson ($M_{\rho}^{(s+u)} \approx 30\%$ and $M_{\rho}^{(t)} \approx
4.5\%$ of $M^{(in)}$). In
comparison with $M_{\pi}$ 
other contributions are significantly smaller. Nonetheless, they 
influence the cross section strongly through interference.
They all have a common relative phase and add constructively. 
The $\rho$, having an opposite sign  counteracts to balance 
their effects.

The ratios quoted above differ considerably
from those predicted by Horowitz et al.\cite{horowitz94}. 
Particularly, the s and u-channel nucleon pole terms for scalar 
meson  exchanges have opposite signs, thus suppressing 
the $\sigma$ exchange contribution and practically eliminating that 
from the $\delta$ meson. 
We note however, that as in Ref.\cite{horowitz94} 
the nucleon pole term contributions from $\pi$ and $\sigma$ 
exchanges add constructively.

We want to emphasize at this stage that, our transition operator 
$ M^{(in)}$ for the production process, involves 
contributions from connected diagrams (Fig. 3) only, and 
accounts for all relativistic and crossing symmetry effects.
Taking  u-channel contributions only, one obtains
far more important effects than 
with both of the s and u-channels. The lesson to be learned here 
is that many small terms 
added up coherently can  explain  a seemingly large discrepancy,
at a time when other small terms which counteract to balance their 
effects are disregarded. Under these circumstances it seems essential 
to preserve crossing symmetry at all stages of the calculations and
treat all contributions on an equal basis.

It is rather surprising that the $\sigma$ contribution is not as 
important as predicted in Ref.\cite{horowitz94}. Yet, it is to be
indicated that in Ref.\cite{horowitz94}, 
only contributions from negative-energy intermediate states are included 
explicitly to the pion production operator
via a Z-graph, while contributions from  positive-energy
nucleon intermediate state are presumably contained in the distorted 
waves describing the initial and final two-nucleon states. Thus,  
different approximations are used in the calculations of the
direct and Z-graph from either s or u- channels and it is not clear
how a delicate balance between these contributions is maintained
throughout the calculations.

In Fig. 8 we draw predictions for the total $pp \rightarrow pp \pi^0$
cross section as obtained for three different value of the 
$\pi N$ $\sigma$-term. 
The small dashed curve shows results 
obtained without the  $\sigma$-meson pole term included. 
Clearly, the overall contribution from nucleon  pole terms only
does not provide the enhancement required to
resolve discrepancy between the calculated cross section and data.
Albeit, the contribution from an isoscalar-scalar $\sigma$-meson pole
term dominates the production cross section and as shall demonstrate below,
with a $\sigma_{\pi N} (0) = 35 ~MeV$, the $\sigma$ pole term
provides the enhancement required to explain cross section data 
near threshold.

The strong dependence of our predictions on the $\pi N$ 
$\sigma$-term deserve a comment. Though still not very well known, 
it seems more likely that the value of this quantity falls in the
range $\sigma_{\pi N} (0) = 35 -45 ~MeV$\cite{gasser91}.  
In the calculations presented in the present work we have not 
included contributions from $\Delta (1232~ MeV)$ isobar excitations.
It remains still to be verified whether the effects of such term would
bring the calculated cross section to agree with data even with a
larger value of $\sigma_{\pi N} (0)$.

\section{FSI and comparison with data }

In comparison with data, the predicted cross section in Fig. 8 
varies very fast with energy due to phase space factor. 
The primary production amplitude (see Fig. 7) is practically 
constant near threshold and therefore contributes very little to
the energy dependence. To cure this deficiency of our predictions
and allow for comparison with data to be made we must account for
ISI and FSI effects.  In what follows we treat ISI and FSI using 
two different approximations, which 
as we shall demonstrate below both yield similar cross sections.\\

$\underline {Approximation I}$

In this approximation the transition operator $M^{(in)}$
is treated as an effective operator acting on nucleon wave functions.
These are calculated using  a phenomenological NN potential.
In this approximation we neglect interactions of the $\pi^0$ produced 
with the outgoing nucleons. This is a standard and usual procedure in
the literature\cite{koltun66,sato97}, reducing a three-body  
problem into effectively a two-body process. 
To be consistent with the OBE picture applied in the present work, we
use the OBE NN potential of Machleidt\cite{machleidt89}; 
potential parameter set C of his table A.2.
The initial and final two-nucleon radial wave functions are 
calculated in momentum space from half off shell $R$ matrix, 
following exactly the procedure of Ref. \cite{machleidt89}. The 
evaluation of the transition amplitude and cross section is
performed as in Ref.\cite{sato97}. We have verified that the calculated
wave functions reproduce well the known experimental phase
shifts\cite{arndt83}.\\

$\underline {Approximation II}$

Here the S-wave production amplitude is assumed to factorize into a 
primary production amplitude $M^{(in)}$, a P-wave ISI factor and  
and an S-wave FSI factor\cite{hep95},
\begin{eqnarray}
T_{2\rightarrow 3} = \langle \Psi_{el,f}^{(+)(3)}| 
M_{2 \rightarrow 3}^{(in)} | \Psi_{el,i}^{(-) (2)} \rangle 
\approx Z_{33} M_{2 \rightarrow 3}^{(in)} X_{22}~,
\end{eqnarray}
where $\Psi_{el,i}^{(-)}(2)$ and $\Psi_{el,f}^{(+)}(3)$ represent
the initial two-body and final three-body wave functions.
The ISI correction factor is taken to be 
$X_{22} \approx |1 + sin (\delta_{^3P_0}) \exp {i (\delta_{^3P_0})} |$.
Here $\delta_{^3P_0}$ is the pp P-wave phase shift. At threshold of 
the $pp \rightarrow pp \pi^0$ reaction, 
$\delta_{^3P_0} \approx -10^O$ so tha ISI corrections 
to the cross section are bound to $\approx \pm 30\%$. 
For a three body process as in our case,
the FSI correction factor is identified\cite{hep95} with the 
elastic scattering amplitude (on mass shell) for the process
$\pi NN \rightarrow \pi NN$ (three particles in to three particles
out) and has the structure of the Faddeev decomposition of the
t-matrix for $3 \rightarrow 3$ transition. 
An important property of this approximation is that the 
different two body interactions among the out going 
particles contribute coherently. Although the meson-nucleon 
interactions are weak with respect to the NN
interaction, they can still be influential through interference. 
This approximation was discussed in length elsewhere\cite{hep95} 
and we skip further details here.
In the analysis presented below the FSI factor is estimated 
from $\pi$N and NN elastic S-wave scattering phase 
shifts\cite{noyes72,hohler83}. The S-wave
NN phase shift is obtained from the effective range expansion 
which includes Coulomb interaction between 
the two protons. We have used the scattering
length $a_{pp} = -7.82$ fm and an effective range $r_{pp} = 2.7$ fm 
of  Ref.\cite{noyes72}. The S11 and S13 $\pi$N  scattering 
lengths are taken to be $a_1 = 0.173 \ m^{-1}_{\pi}$
and $a_3 = -0.101 \ m^{-1}_{\pi}$, respectively\cite{hohler83}.

To compare the results from the two approximations we draw in 
Fig. 9 the partial cross sections corresponding to production via 
a $\sigma$ meson t pole mechanism. 
For neutral pion production, 
the cross sections calculated using the two approximations 
are practically identical. 
We note though that Approximation I
does not account for final meson-nucleon interactions. 
To see how influential the $\pi$N FSI can be, we draw in 
Fig. 10 the cross section corrected 
for FSI interactions assuming pure I=$\frac {1}{2}$ and 
I=$\frac {3}{2}$ interactions for the outgoing
$\pi^0$N pairs. The energy dependence of the cross section differ
significantly for the two channels. But, when
the $\pi$N interactions are taken in the proper 
isospin combination, their overall contribution to the FSI 
factor almost cancels out, leading to an energy 
dependence practically identical with the
one obtained with charged pp FSI (small dash curve) only. 
We want to stress here that this is an accidental consequence 
of the fact that the $\pi^0$p interaction over the allowed 
I=1/2 and I=3/2 isospin channels averages to zero, making 
contributions from diagrams with $\pi^0$p interactions 
ineffective. This may not be the case for charged pions
and $\eta$ meson production\cite{hep95}.

\section{Summary and conclusions }
We have calculated S-wave pion production in $pp \rightarrow pp \pi^0$
using a covariant OBE model, where a boson B created on one
of the incoming protons is converted into a neutral pion on the 
second. The amplitude for the conversion process 
$B N \rightarrow \pi^0 N$ are taken to be the sum of s and u 
pole terms and when allowed a meson pole term in a t-channel.
To  be consistent with the OBE picture of the NN interaction we have 
considered contributions from all of 
the $\pi$, $\eta$, $\sigma$, $\rho$, $\omega$ and $\delta$ 
mesons.  Both the scale and energy dependence 
of the cross section are reproduced rather well. 

Based on a covariant quantum field theory, the transition
operator must be fully covariant and as such it includes
contributions from irreducible connected diagrams only,
each involving large momentum transfer. 
The operator $M^{(in)}$, Eqn. 48, and likewise the calculated
amplitude preserve crossing symmetry and relativistic covariance
throughout the calculations.
Furthermore, terms which involve
three meson vertices are written in a general form. As an example,
the Lagrangian corresponding to a $\sigma \pi \pi$  vertex,
in addition to non derivative terms, includes derivative 
$\sigma ({\partial \pi}{\partial \pi})$ and  
$({\partial \sigma} {\partial \pi}) \pi$ interaction terms. 

The off-shell properties of the various partial production
amplitudes are taken into
account explicitly. For vector and scalar mesons, offshellness is
contained in the nucleon form factor and has a marginal influence.
A pseudovector coupling for the pseudoscalar particles, as assumed
in this work, introduces an extra factor of $q$ to the amplitude.
Consequently for the $\pi$ and $\eta$, each of the s and u nucleon   
pole terms at $q^2 = -3.3 fm^{-2}$ becomes  a factor of $\approx 7$ 
higher in comparison with its value on mass shell. To account for
the off mass shell behavior of the $\sigma$-meson pole
term, we have used Adler's consistency conditions and the isospin
even $\pi^0$p
scattering length to write the $\sigma \pi \pi$ vertex in a rather
general form. Taking all of the s, u and t-channel contributions
into account, we have found that the $\pi^0 p \rightarrow \pi^0p$ 
conversion amplitude off the mass shell is more than a factor
$\approx 15$ higher in comparison with its on mass shell value.      
Consequently, the $\pi$ exchange contribution to the  
amplitude for the $pp \rightarrow pp \pi^0$
reaction dominates the production process and
brings the calculated cross section to agree with data.

We have found that the contribution from a $\sigma$-meson pole 
in a t-channel dominates pion production at threshold.
A term as such accounts effectively for two-pion exchange contributions.  
The $\sigma \pi \pi$ vertex used to calculate this contribution 
depends rather strongly on the $\pi$N $\sigma$-term. However, 
even if we use a value, $\sigma_{\pi N} (0) = 25 ~MeV$, which is 
unrealistically too low, a t-channel $\sigma$ pole term amounts 
to more than a factor of two higher a contribution than that 
of the s and u-channel nucleon poles.
It seems quite impossible to disregard such an important contribution
to the production process. 
Neither the overall contribution from all meson 
exchanges, nor any of the various meson exchange contributions, from 
s and u nucleon pole terms only, can reproduce data for this 
process. In fact,
various contributions of a similar size are found to interfere
destructively and including all of these in a consistent manner,
yields a production amplitude  far 
below what would be  required to explain data. This conclusions seems
unavoidable and independent on the approximation used to account for
ISI and FSI corrections.

A meson production in NN collisions necessarily involves large 
momentum transfer. Thus, two-pion exchanges which describe short 
range interactions are expected to play an important role. In a 
traditional covariant OBE model such contributions are represented
by an effective scalar $\sigma$-meson pole term. In  $\chi$PT 
calculations these should result from expanding the 
existing calculations to 
include one loop two-pion diagrams. Recently, we have extended 
the existing  $\chi$PT calculations\cite{park96,cohen96,sato97}
taking into account all tree and loop diagrams up to chiral
order D=2. With these one loop contributions added, it 
is found that in analogy with a dominant 
$\sigma$ meson pole term, there are substantial
contributions from isoscalar-scalar two-pion t-channel 
exchanges\cite{gedalin98}. 

In summary, two-pion exchanges play an important role in the production
of $\pi^0$ in $pp \rightarrow \pi^0 pp$. By taking these into account,
we have demonstrated that a covariant OBE model can explain 
the existing near threshold neutral pion production data .

\vspace{1.5 cm}
{\bf Acknowledgments} This work was supported in part 
by the Israel Ministry Of Absorption.
We are indebted to  Z. Melamed
for assistance in computation.

\newpage
\begin{figure}
\includegraphics[scale=0.8]{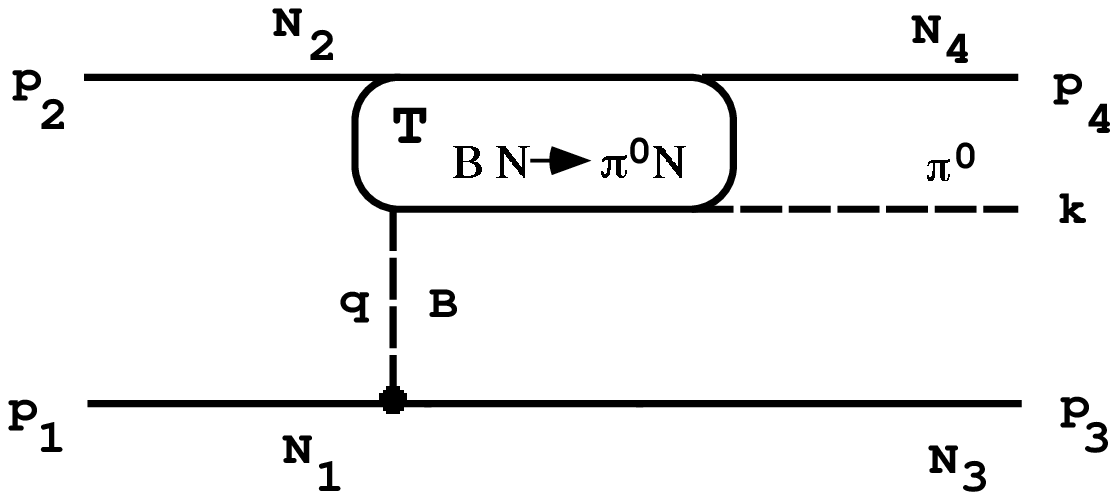}
\caption{The primary production mechanism for the 
$NN \rightarrow NN \pi^0$ reaction. In this figure and following 
figures, solid lines represent nucleons and broken lines mesons.}
\end{figure}

\newpage
\begin{figure}
\includegraphics[scale=0.8]{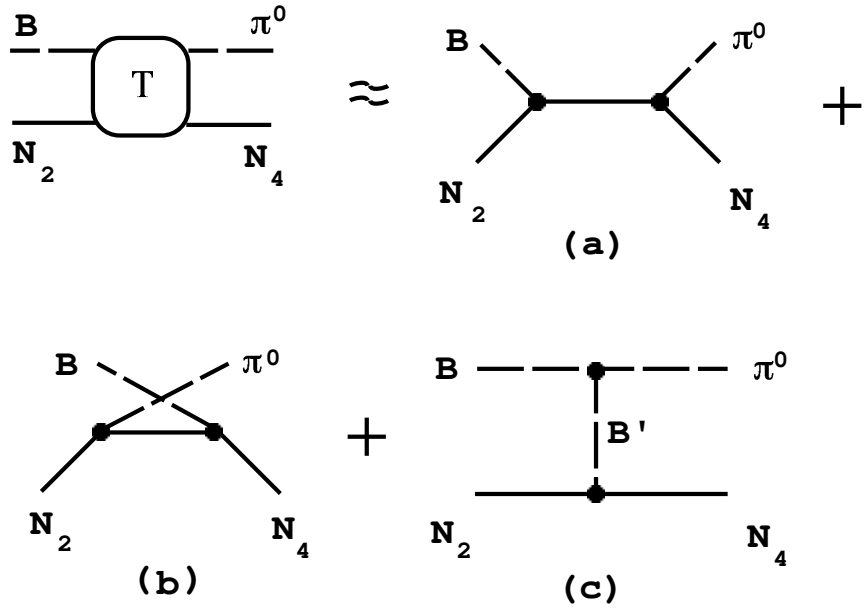}	
\caption{Graphs contributing to the amplitude of the conversion 
process $B N \rightarrow \pi^0 N$. Graphs a and b 
stand for nucleon pole in s and u-channel; the graph c represents 
a B' meson pole in a t-channel.}
\end{figure}
\newpage
\begin{figure}
\includegraphics[scale=0.7]{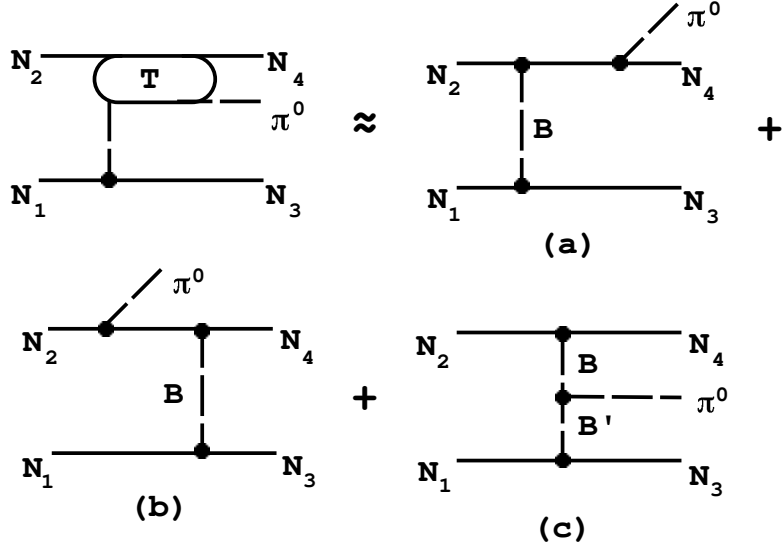}	
\caption{The production amplitude. The mechanisms of graphs a and
b represent pion production on external nucleon lines. Graph c
represents production on an internal meson line. From parity and 
isospin conservations the latter mechanism is limited to production
from $\sigma$ and $\delta$ meson lines only.} 
\end{figure}
\newpage
\begin{figure}
\includegraphics[scale=0.7]{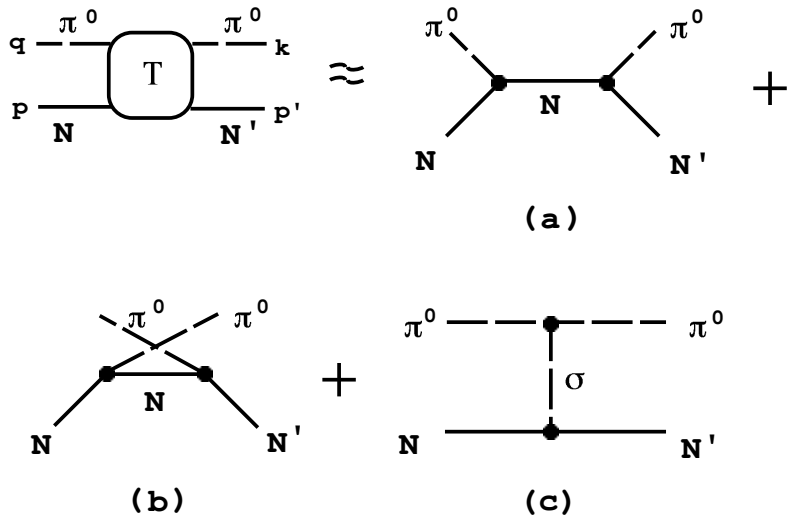}	
\caption{Terms contributing to the $\pi^0 p \rightarrow \pi^0 p$
conversion amplitude. Graph c describes an effective isoscalar 
$\sigma$-meson pole in a t-channel.}
\end{figure}
\newpage
\begin{figure}
\includegraphics[scale=0.7]{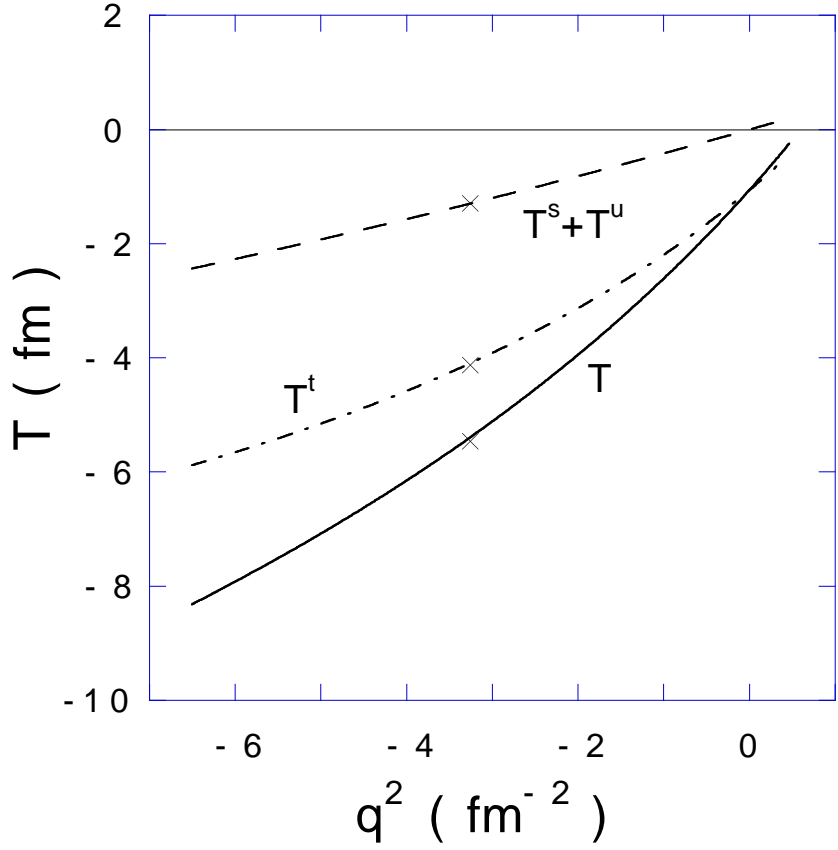}	
\caption{ The off mass shell behavior of the $T_{\pi^0 p 
\rightarrow \pi^0 p}$ amplitude. The contribution from nucleon
s and u-channel pole terms and that from a $\sigma$-meson pole are
drawn as dashed and dot-dashed curves, respectively. The crosses
point threshold of the $pp \rightarrow pp \pi^0$}
\end{figure}
\newpage
\begin{figure}
\includegraphics[scale=0.6]{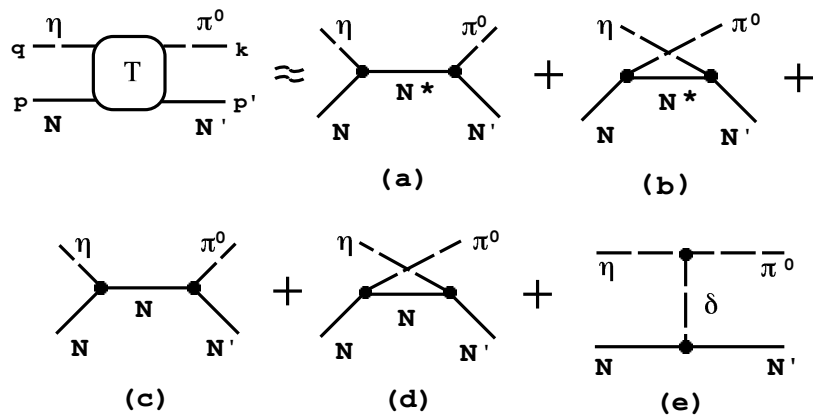}	
\caption{Terms contributing to the conversion amplitude for the
$\eta p \rightarrow \pi^0 p$. Because of the 
strong coupling between the $\eta$-meson and the $N^*$ (1535 MeV) 
nucleon isobar one should allows for contributions from  nucleon 
and nucleon isobar intermediate states.
Graph e describes $\delta$-meson pole in a t-channel.}
\end{figure}
\newpage
\begin{figure}
\includegraphics[scale=0.8]{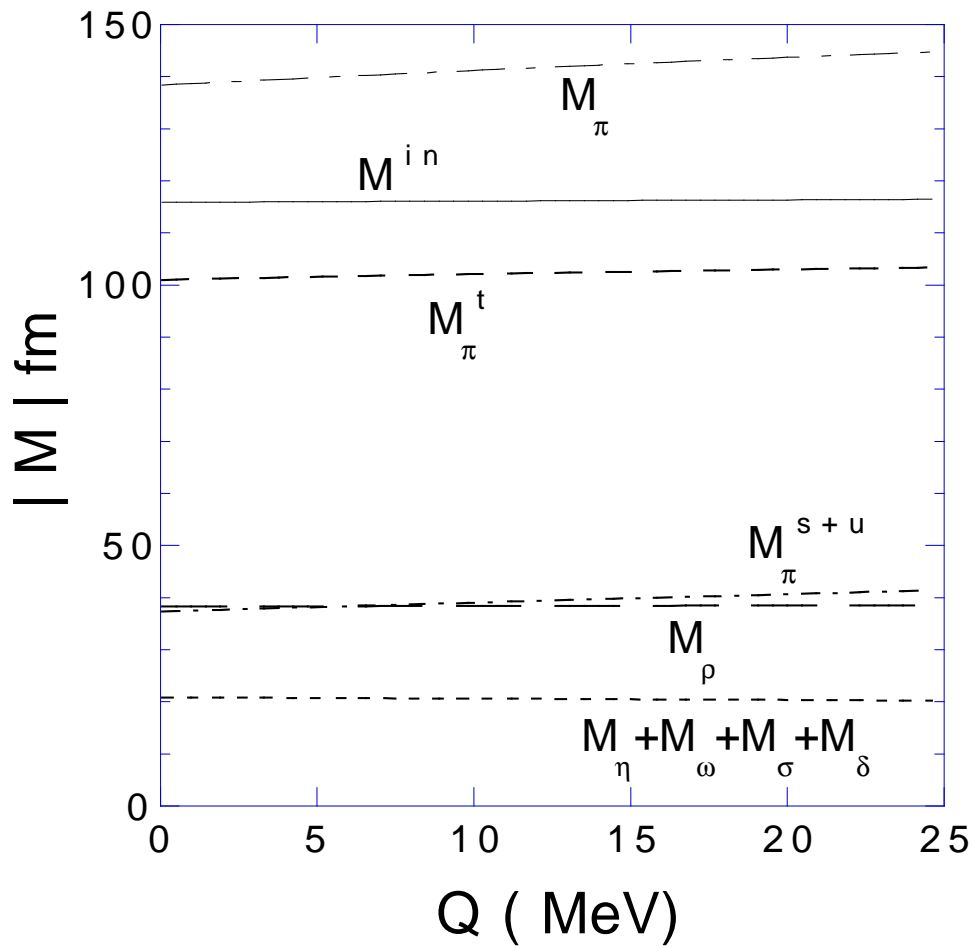}	
\caption{The primary production amplitude for the $pp \rightarrow pp \pi^0$ 
reaction vs. $Q$, the energy available in the overall center 
of mass system. 
Different exchange contributions to $M^{(in)}$ are drawn  
separately. 
}
\end{figure}

\newpage
\begin{figure}
\includegraphics[scale=0.8]{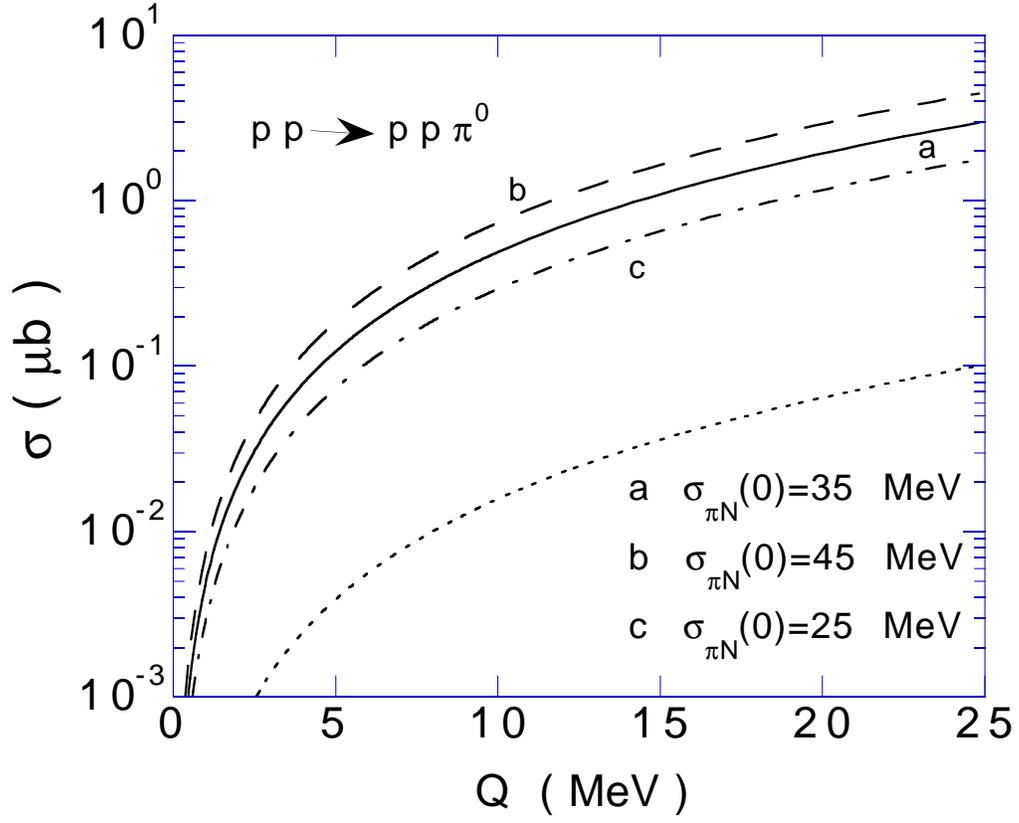}	
\caption{Predictions for the total cross section 
of the $pp \rightarrow pp \pi^0$  reaction for 
different values of $\sigma_{\pi N} (0)$. Predictions with
s and u nucleon pole terms only are shown by the dotted curve.  
All curves are not corrected for FSI.}
\end{figure}

\begin{figure}
\includegraphics[scale=0.7]{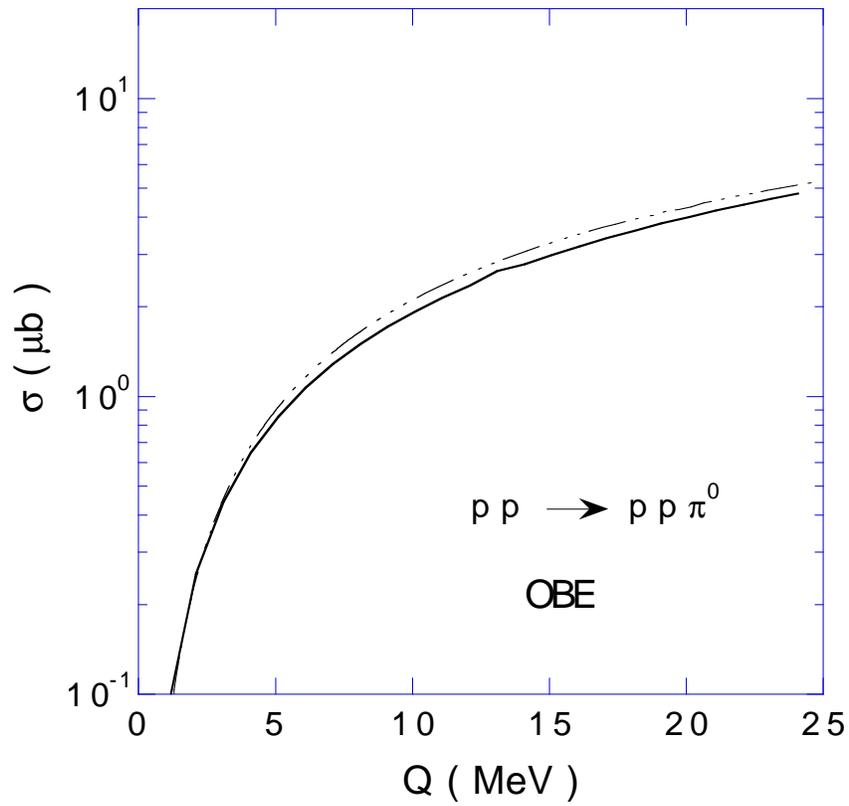}
\caption{ISI and FSI corrections. 
Energy integrated cross sections assuming production via a $\sigma$
meson t pole mechanism only with Approximation I (solid line) and
Approximation II (dot-dashed line)}
\end{figure}

\newpage
\begin{figure}
\includegraphics[scale=0.7]{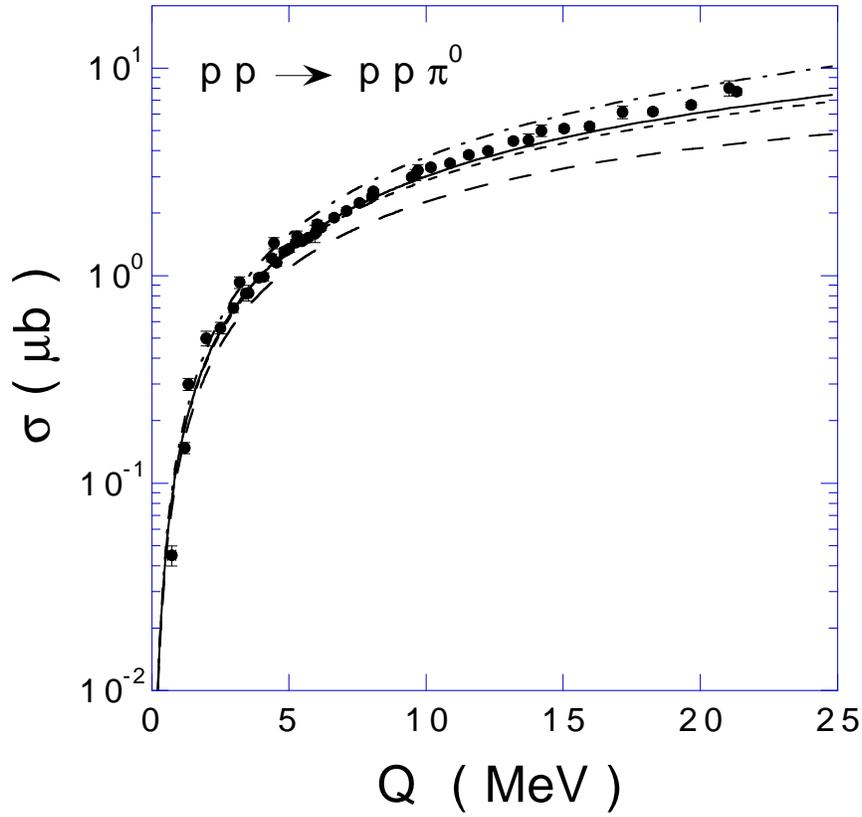}	
\caption{FSI corrections.
Integrated energy cross sections calculated with the assumption that 
the interacting $\pi$N pair is scattered in isospin I=$\frac {1}{2}$
(large dashed curve) and I =$\frac {3}{2}$ (dot-dashed curve). 
The solid line is that obtained with the $\pi$ N interactions
taken in the appropriate isospin combinations. Predictions which account
for the pp FSI only (small dashed curve) are nearly
identical with the solid curve. All predictions are calculated with
$\sigma_{\pi N} (0) = 35 $ MeV.
The data points are taken from Refs.\protect\cite{bondar95,meyer92}}
\end{figure}
\end{document}